\documentclass[3p,times]{elsarticle}

\usepackage{ecrc}

\usepackage{epstopdf}

\volume{00}

\firstpage{1}

\journalname{Nuclear Physics A}

\runauth{J. Casalderrey-Solana et al.}

\jid{nupha}

\jnltitlelogo{Nuclear Physics A}

\usepackage{graphicx}
\usepackage{amsmath,amssymb}

\begin{document}

\begin{frontmatter}



\title{Jet quenching within a hybrid strong/weak coupling approach}



\author[a]{Jorge Casalderrey-Solana}
\author[b]{Doga Can Gulhan}
\author[c,d]{Jos\'e Guilherme Milhano}
\author[a]{Daniel Pablos}
\author[b]{Krishna Rajagopal}

\address[a]{
Departament d'Estructura i Constituents
de la Mat\`eria and Institut de Ci\`encies del Cosmos (ICCUB),
Universitat de Barcelona, Mart\'\i \ i Franqu\`es 1, 08028 Barcelona, Spain
}

\address[b]{Laboratory for Nuclear Science and Department of Physics, Massachusetts Institute of Technology, Cambridge, MA 02139, USA}

\address[c]{CENTRA, Instituto Superior T\'ecnico, Universidade de Lisboa, Av. Rovisco Pais, P-1049-001 Lisboa, Portugal}
\address[d]{Physics Department, Theory Unit, CERN, CH-1211 Gen\`eve 23, Switzerland}

\begin{abstract}
We propose a novel hybrid model for jet quenching, including both strong and weak coupling physics where
each seems appropriate. Branching in the parton shower is assumed to be perturbative and described by DGLAP evolution, while interactions with the medium result in each parton in the shower losing energy as 
at strong coupling, as realized holographically.
The medium-modified parton shower is embedded into a hydrodynamic evolution of hot QCD plasma and confronted with 
LHC jet data.
\end{abstract}


\end{frontmatter}



\section{Introduction}
\label{intro}
We describe a model for the energy loss of jets traversing a strongly coupled 
plasma~\cite{Casalderrey-Solana:2014bpa}. 
This model relies on the separation of scales involved in the different physics regimes of relevance for the quenching dynamics.
 Since high energy jets are produced at a scale $Q\gg \Lambda_{QCD}$, the spectrum is assumed to be under good theoretical control. The consequent parton evolution consisting in the relaxation of this high virtuality parton through successive splittings is described by DGLAP equations. In between the splittings, the partons 
 in the shower can interact with the medium, a plasma at a temperature $T\gtrsim \Lambda_{QCD}$ that is not high enough to ignore strong-coupling effects. 
 The soft in-medium dynamics  are modelled using  insights obtained via gauge/gravity duality.

The energy loss rate of probes passing through a strongly coupled medium has been determined for several non-Abelian theories that have an holographic dual in terms of a gravitational description, to date not including QCD. For these theories, it is not yet possible to treat the hard splittings within the dual gravitational description, which means that the present lack of a holistic explanation motivates a phenomenological approach. We embed the jets from our hybrid model into an expanding hot QCD fluid and compare to LHC data for jet observables.  For a detailed description of this hybrid approach see \cite{Casalderrey-Solana:2014bpa}.

\section{A Hybrid Model}
 
 We use PYTHIA \cite{Sjostrand:2006za} to generate hard processes at such high virtuality that changes in nuclear parton distribution functions can be neglected. Since the in-medium hadronization process is not under good theoretical control,  we stay at the parton level.
For the same reason, we will restrict ourselves to fully reconstructed high energy jet observables, which have limited sensitivity to hadronization effects. 

Each parton travelling through the plasma will be given a formation time ($\tau_f=2E/Q^2$), determining in this way the space-time structure of the complete shower. The expanding plasma is described by a hydrodynamic model whose realistic equation of state \cite{Hirano:2010je} does not have a precise critical temperature, the point below which the plasma is no longer strongly coupled and energy subtraction stops. Accordingly,  the range $180<T_c<200$ MeV is taken to gauge our theoretical uncertainties. In our current implementation, we ignore quenching in the hadronic phase.

\begin{figure}[t]
\centering 
\begin{tabular}{cc}
\includegraphics*[width=.565\textwidth]{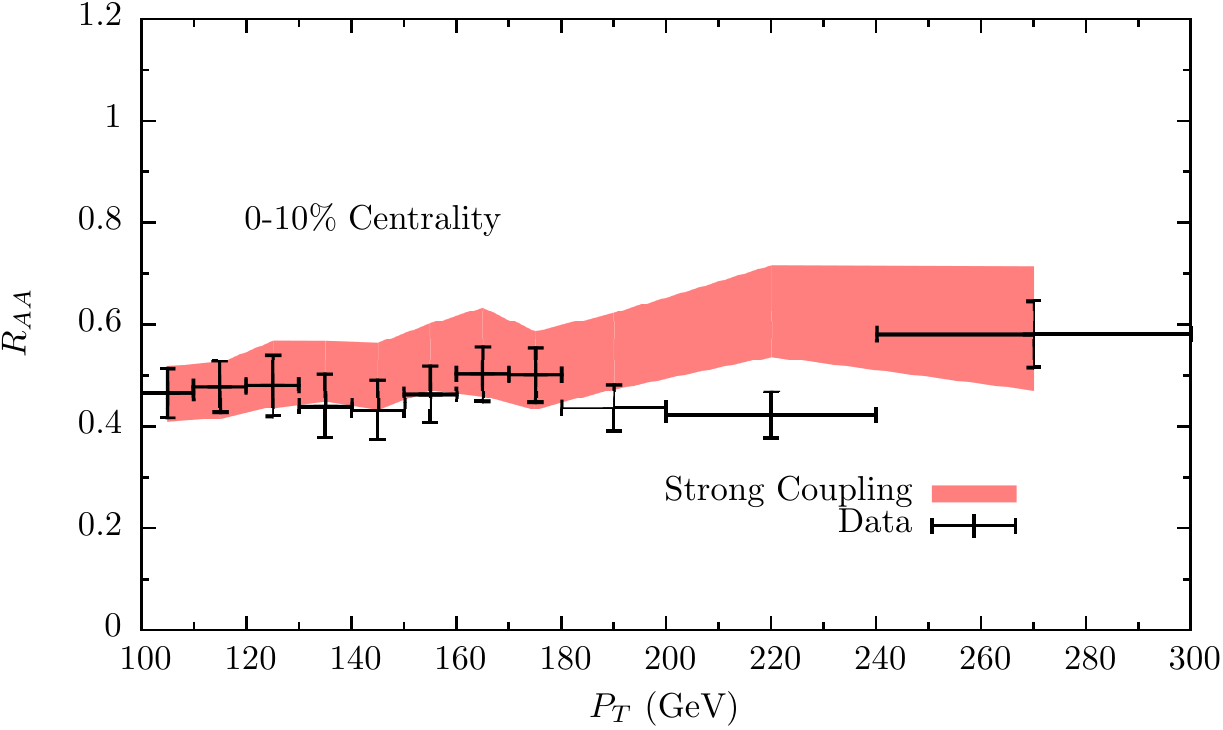}
&
\includegraphics*[width=.395\textwidth]{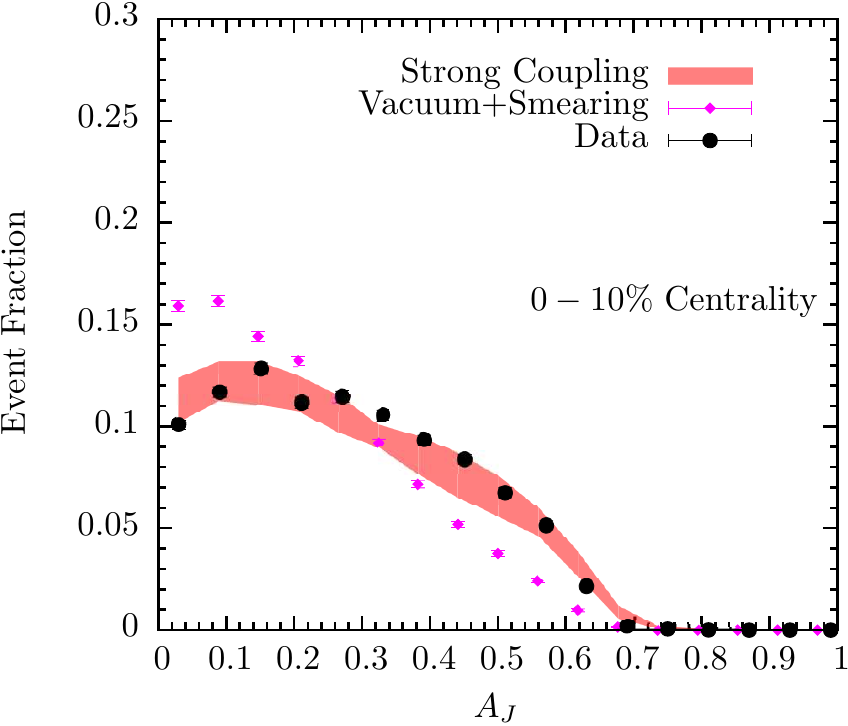}
\end{tabular}
\caption{Left: Jet $R_{AA}$ for central events. The single free parameter in our hybrid model is fitted to the 
left-most data point in this panel. Data from Ref.~\cite{RAACMS}. Right: Dijet asymmetry for central events. The band is the prediction of our model, after its parameter has been fixed as in the left panel. We obtain
a good description of the data, taken from Ref.~\cite{Chatrchyan:2012nia}.  (Systematic error not shown.)}
\label{fig1} 
\end{figure}

To every parton in the medium, we apply an energy loss rate  based on the results of \cite{Chesler:2014jva}, where this quantity was calculated for a light quark traversing a slab of ${\cal N}=4$ supersymmetric Yang-Mills plasma
\begin{equation}
\label{CR}
\frac{1}{E_{in}}\frac{dE}{dx}=-\frac{4x^2}{\pi \,x_{stop}^2 \sqrt{x_{stop}^2-x^2}} \,,
\end{equation}
$E_{in}$ being the initial energy of the parton at its production point and $x_{stop}$ its stopping distance. We determine the latter upon assuming the same dependence on energy and temperature as in the holographic calculation:
\begin{equation}
x_{stop}=\frac{1}{2\, \kappa_{SC}}\frac{E_{in}^{1/3}}{T^{4/3}} \,.
\end{equation}
where we have introduced a dimensionless
parameter $\kappa_{SC}$ whose value is not under good theoretical control. The classical string based computation of \cite{Chesler:2014jva} yields $\kappa_{'\rm SC}=1.05 \lambda^{1/6}$, with $\lambda\equiv g^2 N_c$ the
't Hooft coupling. However, wave-packet based computations as in Ref.~\cite{Arnold:2010ir} yield a $\kappa_{\rm SC}$ that is a pure number of order unity, parametrically independent 
of $\lambda$. Furthermore, the purely numerical component of $\kappa_{\rm SC}$ should be less in QCD
than in ${\cal N}=4$ SYM theory, since QCD has fewer degrees of freedom at the same temperature.
For these reasons, we will treat $\kappa_{SC}$ as a dimensionless fitting parameter to be constrained by data. 


No calculation of the rate of  energy loss for gluons at strong coupling, along the lines of that for
quarks in Ref.~\cite{Chesler:2014jva}, has yet been done.  
Nevertheless, since the parametric dependence of the gluon stopping length
is the same as that for quarks, with a gluon with energy $E$ having
the same stopping length as a quark with energy $E/2$ at large $N_c$
because the gluon is represented by two strings~\cite{Gubser:2008as}, 
we model the rate of energy loss for
gluons using Eq.~(\ref{CR}),
but with
%

%
\begin{equation}
x^G_{stop}=x^Q_{stop}\left(\frac{C_F}{C_A}\right)^{1/3} \,,
\end{equation}
where we have assumed that the factor of 2 is the ratio of the Casimirs in the large-$N_c$ limit.

In the present model implementation of our hybrid approach, we assume that the energy lost
by partons in the shower as they traverse the medium becomes soft particles whose direction
of motion is isotropic, and thus uncorrelated with the direction of the jet.  Such particles are
removed by the background subtraction procedures in experimentalists' analyses of the
jet observables that we compare our predictions to,
and we therefore ignore the lost energy.  This assumption could be improved upon in
future, more sophisticated, implementations of the hybrid approach.

To gain control over the sensitivity of the several studied jet observables to the underlying energy loss mechanism, we have also investigated models in which we repeat our setup, but with a rate of energy loss inspired by perturbative calculations which result in two different path length dependences. First, we use
\begin{equation}
\frac{dE}{dx}=-\alpha_R \frac{C_R}{C_F} T^3 x \,,
\end{equation}
which incorporates (to logarithmic accuracy) the path length dependence of radiative energy loss. Note that the computations based on this rate are only intended as a benchmark against which to compare our hybrid strong/weak coupling approach; we do not see them as superseding 
other more involved implementations of radiative energy loss, in particular those which keep track of the 
radiated gluons.
The second benchmark energy loss rate that we use is
\begin{equation}
\frac{dE}{dx}=-\alpha_C \frac{C_R}{C_F} T^2 \,,
\end{equation}
inspired by collisional energy loss and its characteristic (lack of) path-length 
dependence. We treat $\kappa_{C}$ and $\kappa_R$ as parameters in each model, 
fitting them to the same data used to fit $\kappa_{SC}$ in the hybrid strong/weak coupling model. 

\section{Comparison with Data}

As is done in analyses of experimental data,  
we reconstruct jets using the FastJet \cite{Cacciari:2011ma} anti-$k_T$ algorithm with $R=0.3$ for the rapidity range $|y|<2$. The single free parameter per model is constrained by fitting the jet $R_{AA}$ for $100<p_T<110$ GeV in 0-10\% centrality class. In the left panel of Figure~\ref{fig1}, the jet $R_{AA}$ for the strongly coupled model is shown together with preliminary CMS data \cite{RAACMS}. We observe a weakly $p_T$-dependent shape. In the right panel, we show another inclusive jet observable, the dijet asymmetry, for the same centrality. Even though current data \cite{Aad:2010bu,Chatrchyan:2012nia} have not yet been fully unfolded, there is a smearing procedure used by the CMS collaboration \cite{smear} in which the main systematics of resolution effects are taken into account. Once this has been applied to our results, we compare to data and find highly satisfactory agreement.

\begin{figure}[t]
\centering 
\begin{tabular}{cc}
\includegraphics[width=.435\textwidth]{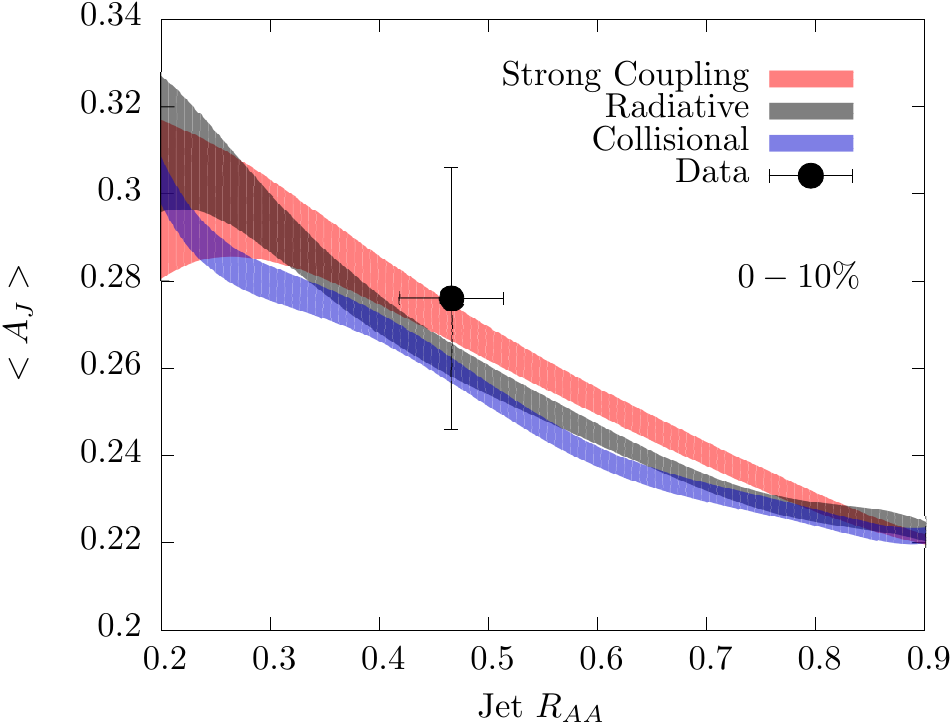}
\includegraphics[width=.545\textwidth]{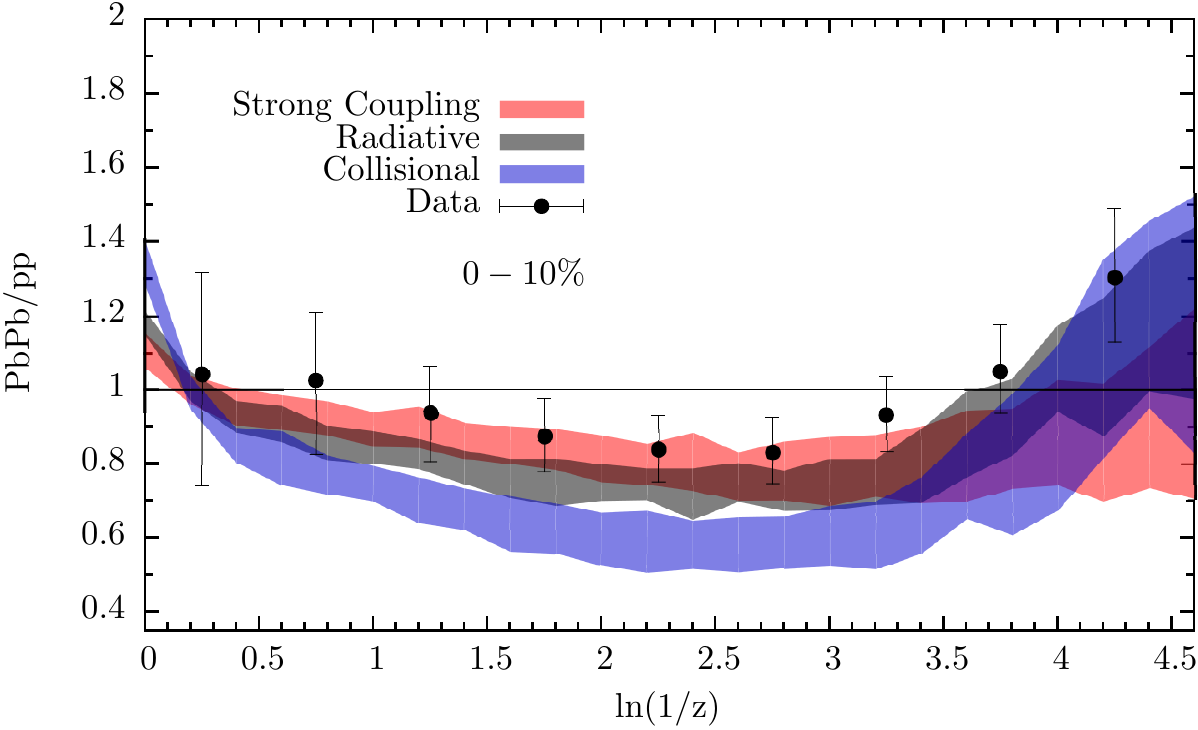}
\end{tabular}
\caption{Left: Scan of  $\langle A_J \rangle$ 
versus $R_{AA}$ for jets with $100~{\rm GeV}<p_T<110~{\rm GeV}$ in central heavy ion collisions at the LHC, as
the free parameter in each model is varied.
Varying the value
of the parameter corresponds to scanning along 
the corresponding colored band. The widths of the band correspond to
the systematic uncertainty coming from varying $T_c$ as described in the text.
Right: Ratio of partonic fragmentation functions for jets with $100~{\rm GeV}<p_T<300~{\rm GeV}$
in central heavy ion collisions to those in pp collisions. 
Data from Ref.~\cite{Chatrchyan:2014ava}.}
\label{fig2}
\end{figure}

\begin{figure}[t]
\centering 
\begin{tabular}{cc}
\includegraphics[width=.40\textwidth]{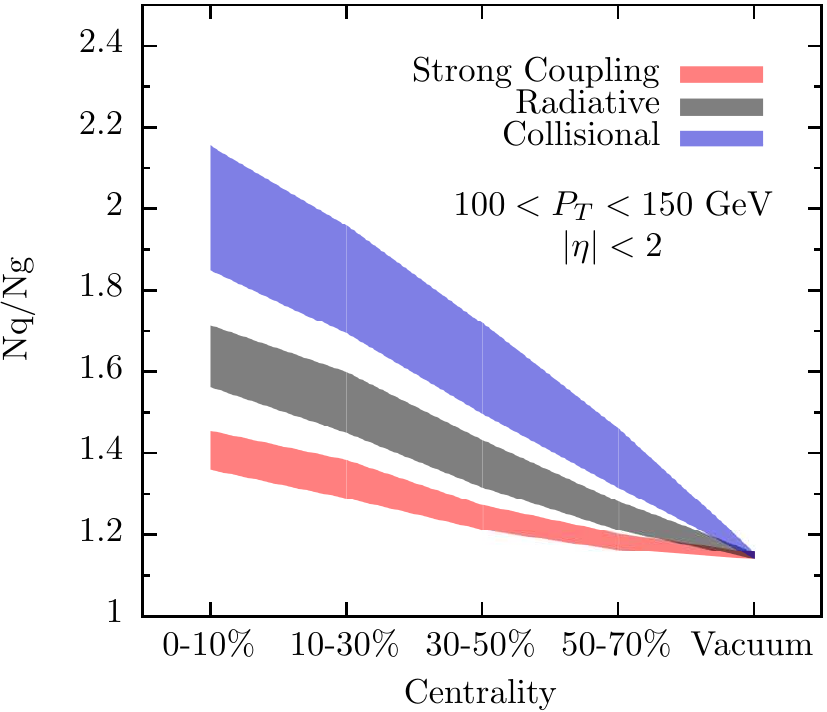}
\includegraphics[width=.585\textwidth]{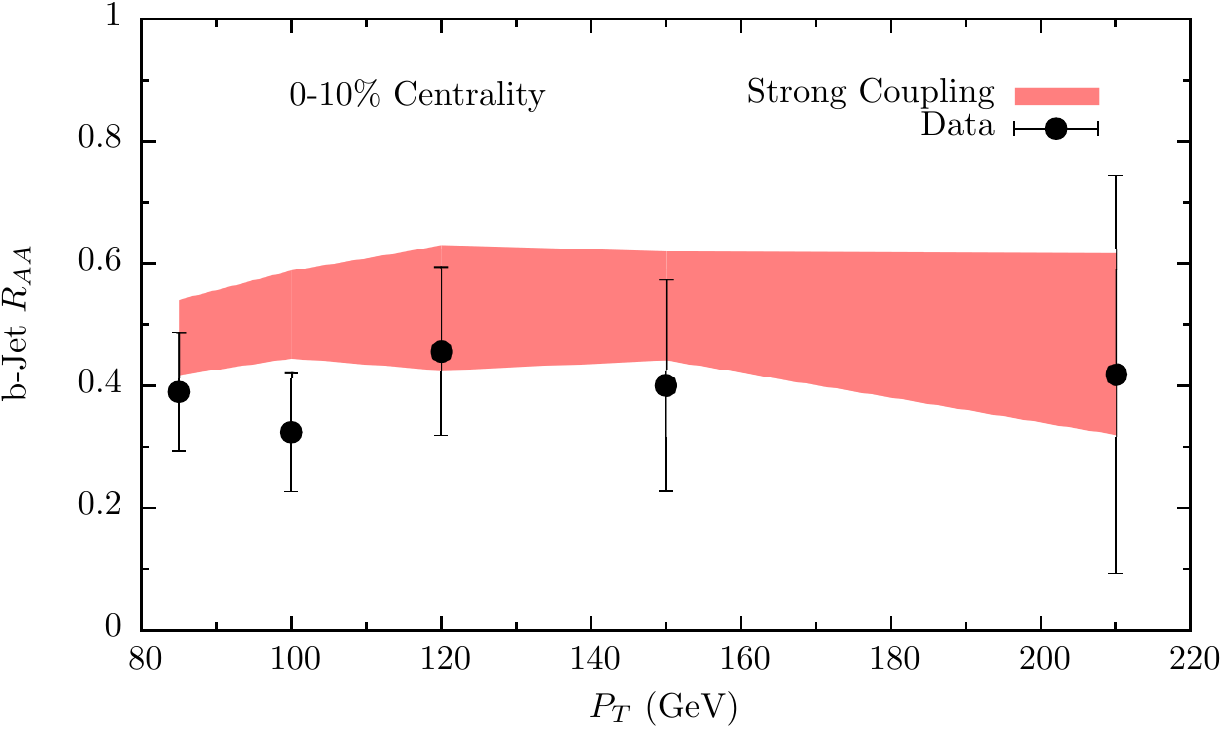}
\end{tabular}
\caption{Left: Number of quark initiated jets over number of gluon initiated vs centrality. Right: b-jet $R_{AA}$ for the strongly coupled model compared to CMS data \cite{bCMS}. Note that the light quark energy loss rate Eq.~(\ref{CR}) has been used for b-quarks, as appropriate at sufficiently high energy.}
\label{fig3}
\end{figure}

The previous success is not exclusive to the strongly coupled model. To see this,  in the 
left panel of Figure~\ref{fig2} we show the predictions of our hybrid model and of the
two weakly coupled benchmarks for the mean dijet asymmetry as a function of the 
jet suppression, where for each point a different value of the fitting parameter was used. Each of the values of $T_c$  leads to a different curve, which in turn end up constructing the observed theoretical bands. One can see how, within current systematic uncertainties, data cannot discriminate between the different models. At the same time a quantification of the needed improvement in precision in jet data can be inferred from the mild separation among the models.

In the right panel of Figure~\ref{fig2}, we show the ratio between in medium and vacuum $partonic$ fragmentation functions. 
Although fragmentation functions depend on hadronization, we have checked that the ratio in Figure~\ref{fig2} is relatively insensitive to the virtuality cut-off of the  evolution, which suggests that a direct comparison with data is meaningful.
The different behavior of the models motivates a more critical discussion.
 The collisional model is strongly disfavored, while the predictions of the 
 strongly coupled model go through the data nicely up to the very soft part of the fragmentation functions. This softest region is where the background sits, and uncorrelated fluctuations can easily enhance the presence of very soft particles with respect to jets in vacuum. We have checked with a thermal distribution that this is  the case.  This is also the region where
 corrections to our present assumption that the energy lost by the jet isotropizes may have discernable effects.

We also compute the ratio of the number of quark initiated jets to the number of gluon initiated jets for each model with $100~{\rm GeV}<p_T<150~{\rm GeV}$ as a function of centrality. This is shown in the left panel of Figure~\ref{fig3}. In contrast to the perturbative models, in the strongly coupled model the ratio of the Casimirs arises suppressed by a power $1/3$, meaning that there is less difference between quark and gluon energy loss at strong coupling than
at weak coupling. Although radiative and collisional energy loss
have the same color charge dependence, the fact that gluon jets are fatter and softer makes them get more suppressed in the collisional model, which is very effective at quenching soft particles. This indicates that more differential 
color-charge-dependent observables may be more sensitive to the underlying properties of the medium.

Motivated by the discussion above, in the right panel in Figure~\ref{fig3} we show a first attempt to describe b-jet suppression within the hybrid approach. At strong coupling, low energy b-quarks lose energy 
via drag~\cite{Herzog:2006gh,Gubser:2006bz,CasalderreySolana:2006rq},
not as in Eq.~(\ref{CR}).
However, at sufficiently high energy, the energy loss of heavy and light quarks should be indistinguishable
and Eq.~(\ref{CR}) will apply to b-quarks. This transition has 
not yet been studied quantitatively, even within the duality. 
Nevertheless, in our first attempt we will assume that the b-quarks are sufficiently
energetic that their  rate of energy loss is the same as that for light quarks. With this additional 
assumption, the strongly coupled model is again compatible with data. Note that, 
unlike for  the other observables in these proceedings, the uncertainty band of our b-quark 
computations is at present sensitive to the statistical error in our set of Monte Carlo events. 
For this reason, we are currently unable  to test the power of this observable to discriminate
between models, which we leave for future work. 

\section{Conclusions}
The simple approach explored here provides a satisfactory agreement between the hybrid strongly coupled model and jet data. The extracted values of the free parameter $\kappa_{SC}$, $0.3 < \kappa_{SC} < 0.4$ are of order one, consistent with holographic calculations and indicating that the stopping length in strongly coupled QCD plasma is about 3 to 4 times longer than that in strongly coupled ${\cal N}=4$ SYM plasma with the same
temperature, a result that is not unreasonable given that QCD has fewer degrees of freedom. Inclusive observables have limited sensitivity to the underlying energy loss mechanism, while others such as fragmentation function ratios can better discriminate, supporting strongly coupled energy loss. Color-charge-dependent observables can also provide a further discrimination, given the different ways in which each model acts on gluon jets relative to quark jets.

\vspace{0.5cm}
\noindent{\bf Acknowledgements}
This contribution to these proceedings corresponds to two talks.  The first talk, given by KR, reported
on his work with P. Chesler 
in Ref.~\cite{Chesler:2014jva}, including in particular the holographic calculations behind 
Eq.~(\ref{CR}). As in the second talk, given by DP, this contribution
to these proceedings is to a significant degree built upon Eq.~(\ref{CR}).
For the derivation of Eq.~(\ref{CR}), as well as for further context and for thoughts about its other 
potential uses, see Ref.~\cite{Chesler:2014jva}.
JCS is a Ram\'on y Cajal fellow. JCS and DP are supported by the grants FP7-PEOPLE-2012-
GIG-333786, FPA2013-40360-ERC, FPA2010-20807, 2009SGR502 and by the Consolider CPAN project. JGM is supported by Funda\c{c}\~{a}o para a Ci\^{e}ncia e a Tecnologia (Portugal) under  project  CERN/FP/123596/2011 and contract 'Investigador FCT - Development Grant'. The work of DCG and KR was supported by the U.S. Department of Energy under grant DE-SC0011090.





\bibliographystyle{elsarticle-num}
\bibliography{bibtex}

\begin{thebibliography}{10}
\expandafter\ifx\csname url\endcsname\relax
  \def\url#1{\texttt{#1}}\fi
\expandafter\ifx\csname urlprefix\endcsname\relax\def\urlprefix{URL }\fi
\expandafter\ifx\csname href\endcsname\relax
  \def\href#1#2{#2} \def\path#1{#1}\fi

\bibitem{Casalderrey-Solana:2014bpa}
J.~Casalderrey-Solana, D.~C. Gulhan, J.~G. Milhano, D.~Pablos, K.~Rajagopal, {A
  Hybrid Strong/Weak Coupling Approach to Jet Quenching }\href
  {http://arxiv.org/abs/1405.3864} {\path{arXiv:1405.3864}}.

\bibitem{Sjostrand:2006za}
T.~Sjostrand, S.~Mrenna, P.~Z. Skands, JHEP 0605 (2006) 026.
\newblock \href {http://arxiv.org/abs/hep-ph/0603175}
  {\path{arXiv:hep-ph/0603175}}.

\bibitem{Hirano:2010je}
T.~Hirano, P.~Huovinen, Y.~Nara, Phys.Rev. C84 (2011) 011901.
\newblock \href {http://arxiv.org/abs/1012.3955} {\path{arXiv:1012.3955}}.

\bibitem{RAACMS}
E.~Appelt for~the CMS~collaboration, these proceedings.

\bibitem{Chatrchyan:2012nia}
S.~Chatrchyan, et~al., Phys.Lett. B712 (2012) 176--197.
\newblock \href {http://arxiv.org/abs/1202.5022} {\path{arXiv:1202.5022}}.

\bibitem{Chesler:2014jva}
P.~M. Chesler, K.~Rajagopal, Phys.Rev. D90 (2014) 025033.
\newblock \href {http://arxiv.org/abs/1402.6756} {\path{arXiv:1402.6756}}.

\bibitem{Arnold:2010ir}
P.~Arnold, D.~Vaman, JHEP 1010 (2010) 099.
\newblock \href {http://arxiv.org/abs/1008.4023} {\path{arXiv:1008.4023}}.

\bibitem{Gubser:2008as}
S.~S. Gubser, D.~R. Gulotta, S.~S. Pufu, F.~D. Rocha, JHEP 0810 (2008) 052.

\bibitem{Cacciari:2011ma}
M.~Cacciari, G.~P. Salam, G.~Soyez, Eur.Phys.J. C72 (2012) 1896.
\newblock \href {http://arxiv.org/abs/1111.6097} {\path{arXiv:1111.6097}}.

\bibitem{Aad:2010bu}
G.~Aad, et~al., Phys.Rev.Lett. 105 (2010) 252303.
\newblock \href {http://arxiv.org/abs/1011.6182} {\path{arXiv:1011.6182}}.

\bibitem{smear}
Y.~Yilmaz, {Jet quenching in heavy-ion collisions at LHC with CMS detector},
  PhD thesis.

\bibitem{Chatrchyan:2014ava}
S.~Chatrchyan, et~al., {Measurement of jet fragmentation in PbPb and pp
  collisions at $\sqrt{s_{NN}}$ = 2.76 TeV}.\href
  {http://arxiv.org/abs/1406.0932} {\path{arXiv:1406.0932}}.

\bibitem{bCMS}
K.~Jung for~the CMS~collaboration, these proceedings.

\bibitem{Herzog:2006gh}
C.~Herzog, A.~Karch, P.~Kovtun, C.~Kozcaz, L.~Yaffe, JHEP 0607 (2006) 013.
\newblock \href {http://arxiv.org/abs/hep-th/0605158}
  {\path{arXiv:hep-th/0605158}}.

\bibitem{Gubser:2006bz}
S.~S. Gubser, Phys.Rev. D74 (2006) 126005.
\newblock \href {http://arxiv.org/abs/hep-th/0605182}
  {\path{arXiv:hep-th/0605182}}.

\bibitem{CasalderreySolana:2006rq}
J.~Casalderrey-Solana, D.~Teaney, Phys.Rev. D74 (2006) 085012.
\newblock \href {http://arxiv.org/abs/hep-ph/0605199}
  {\path{arXiv:hep-ph/0605199}}.

\end{thebibliography}







\end{document}